\documentclass[aip,rsi,reprint]{revtex4-2}

\usepackage{graphicx}
\usepackage{bm}

\usepackage[utf8]{inputenc}
\usepackage[T1]{fontenc}
\usepackage{mathptmx}
\usepackage{etoolbox}

\usepackage[detect-weight=true, detect-family=true]{siunitx}
\DeclareSIUnit{\persqrtHz}{\ensuremath{{\text{Hz}^{-1/2}}}}
\DeclareSIUnit{\sqrtHz}{\ensuremath{\sqrt{\text{\hertz}}}}

\usepackage{subcaption}

\makeatletter
\def\@email#1#2{%
 \endgroup
 \patchcmd{\titleblock@produce}
  {\frontmatter@RRAPformat}
  {\frontmatter@RRAPformat{\produce@RRAP{*#1\href{mailto:#2}{#2}}}\frontmatter@RRAPformat}
  {}{}
}%
\makeatother

\begin{document}


\title{Characterization of non-planar ring oscillators at a wavelength of 1064\,nm for high precision metrology}

\author{H. Vahlbruch}
\email{henning.vahlbruch@aei.mpg.de}
 \affiliation{ 
Max Planck Institute for Gravitational Physics (Albert Einstein Institute), D-30167 Hannover, Germany \\and \\Leibniz University Hannover, D-30167 Hannover, Germany
}
\author{F. Meylahn}
\email{fabian.meylahn@aei.mpg.de}
 \affiliation{ 
Max Planck Institute for Gravitational Physics (Albert Einstein Institute), D-30167 Hannover, Germany \\and \\Leibniz University Hannover, D-30167 Hannover, Germany
}
\author{B. Willke}
 \affiliation{ 
Max Planck Institute for Gravitational Physics (Albert Einstein Institute), D-30167 Hannover, Germany \\and \\Leibniz University Hannover, D-30167 Hannover, Germany
}

\date{\today}

\begin{abstract}


Ultra-stable laser light is essential for high-precision interferometric measurements, in particular for the next generation of gravitational wave detectors, where high power lasers with unprecedented low power and frequency noise are demanded. Since the seed laser for high-power laser system has a large influence on the overall noise characteristics, the use of the lowest noise seed laser is beneficial. This study compares a newly developed seed laser, based on a non-planar ring oscillator (NPRO) design, at a wavelength of \SI{1064}{nm} with two commercial NPROs and shows that the new laser exhibits ten times lower power and frequency noise. This noise advantage is retained even after subsequent amplification to \SI{40}{W}.

\end{abstract}

\maketitle

\section{Introduction}

Since their first development over 40 years ago \cite{Kane1985}, non-planar ring oscillators (NPRO) have proven to be a class of laser sources with exceptional stability. They are suitable for high-precision measurements and are therefore used in a variety of experiments, such as in the field of quantum optics. Particularly, they are a key component of the complex high-power laser systems in the current generation of ground-based gravitational wave detectors \cite{Tse19, Acernese19, Lough21, GeoKagra2022C}.

The monolithic design of NPRO lasers, in which the resonator is formed by the laser-active solid-state material, ensures robustness and stability. The pump light from a laser diode is coupled into the crystal and the generated laser light is extracted through a dielectrically coated surface, while the total internal reflection at the other polished surfaces of the non-planar resonator allows for a simple monolithic construction and minimizes losses. By using the Faraday effect caused by an external magnetic field that penetrates the NPRO in combination with the polarization-dependent dielectric coating of the NPRO crystal, uni-directional single-frequency and single-mode operation is achieved.\cite{Kane1985, Kane1988, Nilsson1989}.

The emission wavelength of NPROs is primarily determined by the choice of laser-active material in combination with a dielectric optical coating. For example, Neodymium (Nd)-doped Yttrium aluminum garnet crystals can emit at \SI{1064}{\nm} or \SI{1319}{\nm}, while Holmium and Erbium doping allows operation at other wavelengths, such as \SI{1645}{\nm} \cite{Ter-Gabrielyan2010, NPRO1645nm} or \SI{2}{\um} \cite{Ebert2025}.
Over the years, numerous advances have been made with regards to the available output power \cite{Freitag1995}, tunability \cite{Kane2025}, miniaturization \cite{Yu2023}, and noise performance \cite{Heurs2006}. Because of the parallel development of external solid-state or fiber-based amplifier stages, which can scale the laser power of an NPRO, the focus of the NPRO development has shifted from maximizing the output power to optimizing the noise behavior, tunability, and miniaturization of NPROs as seed lasers for such amplifier systems. Moreover, recent developments in laser systems have shown that the use of fiber-based phase modulators can provide significant benefits in terms of modulation depth and speed \cite{Spencer2022,Meylahn2022}. However, these modulators typically only handle a few hundred milliwatts of input power, necessitating the use of low-power seed lasers in such laser systems.

Multi-stage laser systems with high output power and particularly low power and frequency noise are relevant for the current and next generations of ground-based gravitational wave detectors and their associated fundamental research. The requirements for laser stability are challenging in this area. Even though sophisticated laser stabilization concepts are in use, a good un-stabilized (free-running) noise performance of the lasers is crucial. Repeated characterizations of laser sources have been conducted to identify the best-suited devices and to further develop the laser systems of these detectors \cite{Kwee2008,Meylahn2021,Bode2020a}. For the next generation of gravitational wave detectors, such as the planned Einstein Telescope (ET) in Europe \cite{ETdesignStudy} and the Cosmic Explorer (CE) in the USA \cite{cosmicExplorer}, the laser noise requirements are even more stringent than for current detectors \cite{Cahillane2021} such that novel lasers with lower noise would be beneficial.

In this article, we characterize and compare NPRO lasers that emit at \SI{1064}{\nm} wavelength, one of which was in-house developed and has substantially lower noise properties than the others. We evaluate the suitability of this newly developed NPRO for current and future gravitational wave detectors operating at this wavelength. Our comparative results are also relevant in other areas, such as quantum optics \cite{mHzSqueezing}, the search new particles, dark matter or quantum gravity signatures \cite{Kozlowski2025,Patra2024}, and \si{nm}-inter-satellite metrology \cite{Abich2019}.

\begin{figure*}[htbp]
\centering
\includegraphics[width=0.7\textwidth]{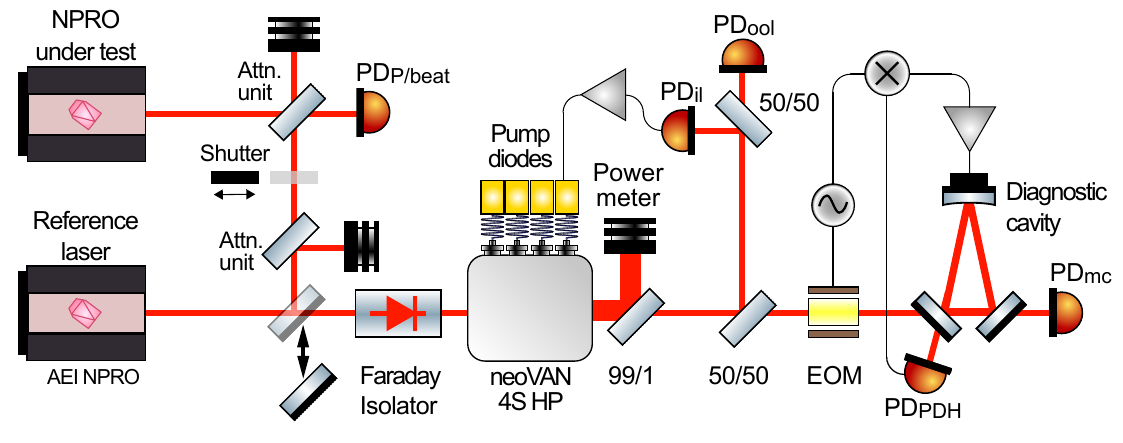} 
\caption{The experimental setup consists of three parts. On the left side, the NPRO lasers under test are located, along with a reference laser for individual NPRO power noise and laser frequency noise measurements. In the middle of the setup, the laser amplification stage is illustrated, where the amplified laser beam is generated. On the right side, the power stabilization and diagnostics setup for the amplified laser beam is shown.}
\label{fig:setup}
\end{figure*}

\section{Setup} 
The experimental setup used to characterize the performance of non-planar ring oscillators is shown in Fig.~\ref{fig:setup}.
The NPRO lasers under test were a Mephisto laser from Coherent Corp. (\SI{500}{\mW} output power) \cite{MephistoDatasheet}, a Lumentum JDP-126M-1064-500 laser from Lumentum Holdings Inc. (\SI{500}{\mW} output power) \cite{LumentumDatasheet}, and the new in-house developed NPRO at the Albert-Einstein-Institute (AEI), Hannover, Germany, which we call in this paper AEI NPRO (up to \SI{400}{\mW} output power). We compared their performance in terms of laser power noise and frequency noise.

To measure the laser power noise, the shutter in Fig.~\ref{fig:setup} was closed and each laser was successively inserted into the 'NPRO under test' location of the setup. The laser output power was attenuated, and the residual laser light was directly detected on an InGaAs photodiode (PD$_{\text{P/beat}}$) with a \SI{2}{\mm} sensor diameter. This photodiode, capable of detecting up to \SI{50}{\mW} of incident laser power, was read out via an active transimpedance amplifier \cite{Kwee2008}. After a warm-up time and thermalization of each laser, the amplitude spectral density of the power measurement was captured, calibrated, and normalized by the total incident power to determine the relative power noise of the NPRO under test.

For the characterization of the laser frequency noise, a second laser (an AEI NPRO) served as a reference. Its output field was interfered on a beam splitter with the output field of the laser under test. By adjusting the crystal temperature of the NPRO under test, the emission laser frequency was matched to the reference laser frequency with a difference of a few tens of \si{\MHz}. The beat signal detected by PD$_{\text{P/beat}}$ was analyzed using a phase meter implemented in a Moku:Lab measurement device by Liquid Instruments. The time series of the phase meter output which corresponds to the difference of the laser's  frequencies was recorded, and its amplitude spectral density was calculated using the Welch method.

To demonstrate power scaling with the newly developed NPRO and to analyze the noise of the amplified laser beam, the experimental setup was extended. The reference laser beam was directed into a single-pass Nd:YVO$_4$ power amplifier \cite{Thies2019,Bode2020a} from neoLASE GmbH, shown in the central part of Fig.~\ref{fig:setup}. A Faraday isolator was inserted to protect the AEI NPRO laser from back-reflected light, for example coming from the power amplifier.
The neoVAN amplifier, which hosts four amplification stages pumped at a wavelength of \SI{878}{\nm} by fiber-coupled laser diodes, was used to amplify the AEI NPRO's laser beam. One of the laser diode power supplies included an integrated current shunt, which served as a power actuator \cite{Thies2019}. The output power of the amplifier was measured using a thermal power meter, except for \SI{1}{\percent} of the laser light which was diverted to the stabilization and characterization part of the experiment. To analyze the power noise, a photodiode identical to PD$_{\text{P/beat}}$ was used (PD$_{\text{ool}}$).

The laser frequency noise of the amplified beam was measured using a diagnostic triangular ring cavity, the length of which was stabilized via the Pound–Drever–Hall technique \cite{Drever1983}. The control loop feedback signal was applied to a piezoelectric element attached to one of the cavity mirrors, allowing for precise control of the cavity length using in-house developed analog electronics. The cavity was shielded against acoustic disturbances with an aluminum enclosure. The error and control signals of the feedback control loop were captured and calibrated to compute the frequency noise, which represents the differential noise between the cavity length fluctuations and the laser frequency noise.

In the final step, an analog feedback control loop was implemented to stabilize the amplified laser power. In the control loop, the laser power is sensed using an in-loop photodiode (PD$_{\text{il}}$) and actuation is applied onto the neoVAN's power actuator \cite{Thies2019}. To verify the effectiveness of the stabilization, the photodiode (PD$_{\text{ool}}$) was used as an independent out-of-loop detector to measure the power noise of the stabilized high-power beam. In addition to the characterization tools shown in Fig.~\ref{fig:setup}, the power at different locations of the experiment were measured using power meters, the spatial beam quality was evaluated by mode-scan measurements using the triangular cavity, and the polarization extinction ratio was determined by measurements after a waveplate and polarization beam splitter combination.

\begin{figure*}[htbp]
    \centering
    \begin{minipage}{0.49\textwidth}
        \centering
        \includegraphics[width=\linewidth]{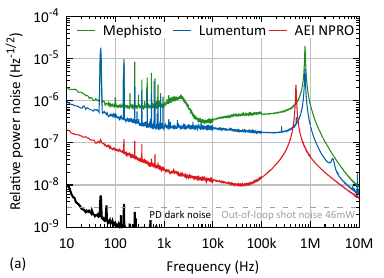}
         \label{fig:rpnFree}
    \end{minipage}
    \hfill
    \begin{minipage}{0.49\textwidth}
        \centering
        \includegraphics[width=\linewidth]{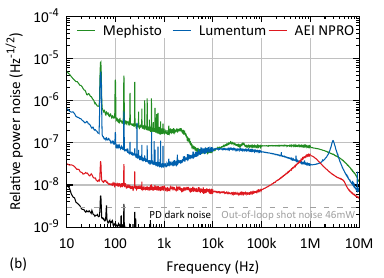}
         \label{fig:rpnNE}
    \end{minipage}
    \caption{A comparison of the relative power noise of three NPRO-based laser sources. In \textbf{(a)} the measurements without laser internal noise reduction are shown, whereas the results with the laser internal noise suppression engaged are presented in \textbf{(b)}. To ensure that the out-of-loop measurements were not limited by photon shot noise or electronic dark noise, an in-house made high-power and low-noise photo detector was used.}
    \label{fig:sidebysideRPN}
\end{figure*}

\section{Results and Discussion}

We first present the characterization of the different non-planar ring oscillators in the following section. We then discuss the results of power amplification of the lowest noise NPRO, namely the AEI NPRO, and illustrate how laser noise propagates through a single-pass laser amplifier system. All noise measurements presented here were obtained using out-of-loop sensors, which are not part of any feedback control systems for the tested lasers.

\subsection{Non-planar ring oscillator characterization}

The relative power noise of the lasers was measured both with and without the lasers' internal noise reduction activated, as shown in Fig.~\ref{fig:sidebysideRPN}a and \ref{fig:sidebysideRPN}b. The free-running noise of the NPRO laser emission shows a characteristic relaxation oscillation in the range of \SI{400}{\kHz} to \SI{1}{\MHz}, as visible in Fig.~\ref{fig:sidebysideRPN}a. The frequency of the relaxation oscillation is influenced by the lasers' output power levels, resonator characteristics, and gain in the laser medium \cite{Harb1997}.

The measured relative power noise of the Mephisto laser is higher than that of the Lumentum laser, while the AEI NPRO laser exhibits the lowest noise. 

\begin{figure}[tbp]
        \centering  
        \includegraphics[width=\linewidth]{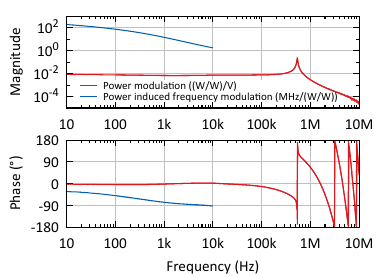}
    \caption{
     The pump-current-to-emission-power transfer function was measured for the AEI NPRO laser, showing a resonant enhancement around the relaxation oscillation and a decay above it (red trace). Additionally, the pump-power-to-emission-frequency transfer function was measured using the beat note between two AEI NPRO lasers and a phase meter (blue trace).}
     \label{fig:powerMod}
\end{figure}

The internal noise reduction is a feedback power stabilization implemented in all three tested NPRO lasers, also called noise-eater. It is realized with an internal control loop: A fraction of the laser output power at \SI{1064}{\nm} wavelength is sensed by an internal sensor, and a servo controller then applies feedback to the current sent to the pump diodes at a wavelength of \SI{808}{\nm}, with the primary focus of reducing power noise, particularly targeting the relaxation oscillation. This control loop is effective due to the resonance peak of the pump-to-output power transfer function (see Fig.~\ref{fig:powerMod}), which enables a high noise suppression factor at the relaxation oscillation frequency \cite{Kane1990}

Fig.~\ref{fig:sidebysideRPN}b shows the relative power noise of the tested lasers with the lasers internal noise reduction engaged. The noise reduction is evident across all lasers, with a significant decrease in noise levels in the range from Fourier frequencies of \SI{10}{\Hz} up to \SI{1}{\MHz}. 
However, the differences in noise levels remain similar, with the Mephisto and Lumentum lasers exhibiting noise levels in the range of \SI{1e-7}{\persqrtHz}, while the AEI NPRO laser achieves a relative power noise level below \SI{1e-8}{\persqrtHz} from \SI{100}{\Hz} up to \SI{100}{\kHz}. Above \SI{1}{\MHz}, the feedback controllers introduce excess noise, limiting the noise suppression achieved by the internal noise reduction.

\begin{figure}[tbp]
        \centering  
        \includegraphics[width=\linewidth]{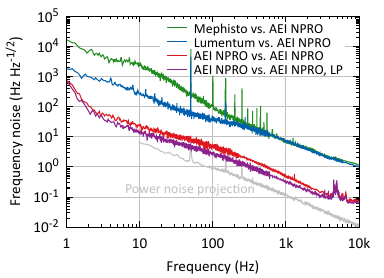}
         \caption{Frequency noise of the NPRO lasers derived from beat note measurements, with all measurements  taken with the internal noise reduction of the lasers engaged. Notably, the AEI NPRO laser achieved an even lower frequency noise level when operated at \SI{250}{\mW} output power (LP).}
         \label{fig:frq}
\end{figure}

The coupling between pump power fluctuations and laser frequency noise in NPRO lasers is a well-known process that significantly impacts the overall frequency noise performance. 
This coupling arises from fluctuations in the pump power, which heat the laser medium, causing changes in the refractive index and thermal expansion. As a result, the optical path length of the NPRO is altered, leading to a change in the resonance frequency and, consequently, the emitted laser frequency~\cite{Day1991,Willke2000}.
To quantify the coupling between pump power fluctuations and laser frequency noise, measurements were conducted on the AEI NPRO laser. The analysis involved examining the beat note between two lasers and the modulation input-to-output power transfer function, as shown in Fig.~\ref{fig:powerMod}. 
At frequencies above \SI{200}{\Hz}, the transfer function from relative power changes to the emission frequency exhibits an $1/f$-shape, which is attributed to the local heating of the laser medium by the pump beam. At lower frequencies the transfer function flattens because the thermal diffusion becomes more significant than the thermal low-pass associated to the materials heat capacity.
Power fluctuations of the pump diodes can limit the frequency noise performance of the NPRO lasers by the here described coupling. 
For the AEI NPRO laser, this noise was projected using the transfer function from output power to laser frequency (in Fig.~\ref{fig:powerMod}) and the power noise measurement to estimate the frequency noise, as shown by the grey curve in Fig.~\ref{fig:frq}. 

Fig.~\ref{fig:frq} shows the frequency noise measurements of the lasers obtained via beat frequency measurements between two independent lasers. For the AEI NPRO laser, the measurements were performed using two identical lasers, assuming uncorrelated laser frequency noise between the lasers. This assumption implies that the measured noise is a factor of $\sqrt{2}$ higher than the actual frequency noise of an individual laser. For the Mephisto and Lumentum lasers, one of the AEI NPRO lasers were used as a reference. As this reference has a significant lower frequency noise than the other NPROs under test, the influence on the measured frequency noise of the Mephisto and Lumentum was less than \SI{0.5}{\percent}, for a noise level difference of one magnitude. All measurements were conducted with the internal noise reduction of the lasers engaged.
The AEI NPRO laser's frequency noise level is approximately one magnitude lower than that of the Mephisto or Lumentum lasers, as shown in Fig.~\ref{fig:frq}. This result is consistent with the well-documented coupling between laser power noise and frequency noise in NPRO lasers, as reported in \cite{Day1991,Willke2000,Heurs2004,Huntington2007}.
Since the exact layout and design of the internal optical and electronic components of the commercial products are not known, a direct comparison of the  noise contributions, for example, from pump laser diode or temperature controllers, is not possible. However, with the AEI NPRO, the low noise level was achieved through the careful design of a low-noise current driver, good temperature stabilization, an internal noise suppression system with active control below \SI{10}{\Hz}, and the optimization of the spatial overlap between the pump and signal fields.

The frequency and power noise levels measured for the Mephisto and Lumentum lasers are consistent with results in the respective data sheets~\cite{MephistoDatasheet,LumentumDatasheet}, and a comprehensive characterization of the \SI{2}{\W} Mephisto version as provided in~\cite{Kwee2008}.

\begin{figure}[htbp]
    \centering
        \includegraphics[width=\linewidth]{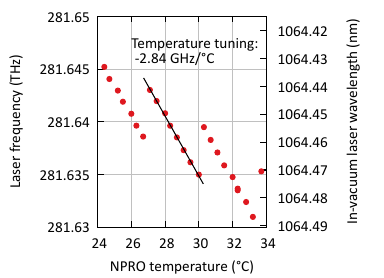}
        \caption{The absolute laser frequency, captured with a wavemeter instrument, and the laser frequency thermal tuning range of the AEI NPRO is illustrated. The characteristic mode hops, where another free spectral range of the NPRO gains the highest cross section with the optical gain, can be observed every \SI{3}{\K}.} 
        \label{fig:thermal}
\end{figure}

To compare the AEI NPRO to the specifications of the other NPROs, we used 
a length-scanned triangular cavity \cite{Kwee2008}, and measured the higher-order spatial mode content of the beam emitted by the AEI NPRO laser to be less than \SI{1.3}{\percent}, corresponding to a $M^2< 1.03$, as the higher-order modes were predominantly third and fourth order modes~\cite{Siegman1990}. 
The AEI NPRO laser frequency tuning via thermal actuation on the NPRO was measured using a WS7-60 wavemeter from HighFinesse GmbH. The measured frequency tuning, which had a bandwidth of \SI{0.8}{\Hz}, exhibited a tuning coefficient of \SI{-2.84}{\GHz\per\K} and a mode-hop-free operation range of approximately \SI{3}{\K}, as shown in Fig.~\ref{fig:thermal}.

\begin{figure}[htbp]
        \centering
        \includegraphics[width=\linewidth]{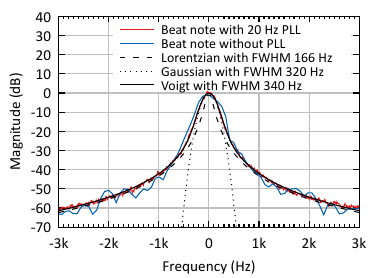}
    \caption{The beat measurements of two AEI NPRO lasers with enabled internal noise reduction are shown, obtained using two different measurement configurations. The free-running beat note was measured with a resolution bandwidth of \SI{62}{\Hz} and no averages, while the beat note with a \SI{20}{\Hz}-PLL was measured with a resolution bandwidth of \SI{18}{\Hz} and 100 averages.}
    \label{fig:beat}
\end{figure}

The linewidth of the AEI NPRO laser was directly measured by analyzing the beat note between two identical AEI NPRO lasers using a spectrum analyzer, as shown in Fig.~\ref{fig:beat}. This measurement was conducted with a resolution bandwidth of \SI{62}{\Hz} and no averages. 
For a measurement with averages, an offset phase lock loop (PLL) was implemented, to compensate for slow frequency drifts. The control loop bandwidth was chosen to be approximately \SI{20}{\Hz}, which is well below the expected laser linewidth, and thus does not influence the measurement result. The resolution bandwidth was \SI{18}{\Hz} and the number of averages 100. 

The measurements indicate a linewidth of significantly less than \SI{1}{\kHz}, with the Gaussian profile having a greater influence on the full width half maximum (FWHM) linewidth than the Lorentzian. The Gaussian profile is related to the $1/f$-shape of the laser frequency noise \cite{Stephan2005}, see Fig.~\ref{fig:frq}, which is caused by the pump-power to laser frequency noise coupling. Assuming uncorrelated noise between the two lasers, the actual linewidth of a single laser would be $\sqrt{2}$ times smaller, resulting in a full width half maximum of \SI{240}{\Hz}.
It should be noted that one of the AEI NPRO lasers was operated for more than \SI{2000}{\hour} without showing any degradation in power or noise performance.

\subsection{Characterization together with a solid-state amplifier}

Several applications in the field of quantum optics and high precision metrology  require laser powers in the higher watt range. To address these high-power requirements, we used the lowest-noise NPRO laser, namely the AEI NPRO, as a seed laser for a four-stage solid-state Nd:YVO$_4$ amplifier from neoLASE GmbH.

\begin{figure}[htbp]
        \centering
        \includegraphics[width=\linewidth]{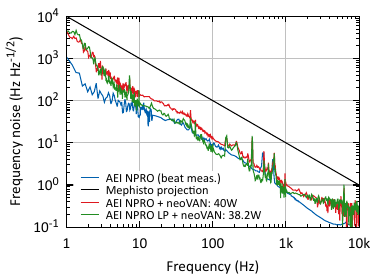}
         \caption{Frequency noise measurements conducted using the diagnostic ring cavity. The frequency noise of the amplified laser beams is compared to the frequency noise of the seed laser used for this measurement and a typical noise projection of a Mephisto NPRO~\cite{Kwee2008} for reference. The noise measurements of the high-power beams are dark-noise-corrected, which affects the data above \SI{1}{\kHz}.
         All shown measurements were performed with engaged internal noise reduction of the AEI NPRO.}
        \label{fig:frqneoVAN}
\end{figure}

For this experiment, as shown in Fig.~\ref{fig:setup}, the AEI NPRO laser is free-space coupled to the amplifier through a Faraday isolator, protecting the seed laser from back reflections. Operating the AEI NPRO laser at an output power of \SI{400}{\mW}, we measured an amplified power of \SI{40.4}{\W} with an amplification factor of approximately 100. The polarization extinction ratio post-amplification was measured to be \SI{30}{\dB}, and the higher-order mode content was less than \SI{2.7}{\percent}.
The frequency noise of the amplified beam was analyzed using a triangular cavity and is presented in Fig~\ref{fig:frqneoVAN}. Consistent with previous investigations of single-pass solid-state amplifiers, no significant difference was observed between the frequency noise of the amplified laser beam and the seed laser. However, it's worth noting that an early version of the AEI NPRO laser was used for this test, which did not achieve the full noise performance demonstrated in Fig.~\ref{fig:sidebysideRPN}b and Fig.~\ref{fig:frq}.
The measured frequency noise closely approaches the (known) stability limits of the triangular mode cleaner, particularly with acoustic coupling peaks between \SI{100}{\Hz} and \SI{1}{\kHz} affecting the diagnostic cavity length \cite{Kwee2009}. 
Reducing the AEI NPRO output power down to \SI{250}{\mW} (low-power mode, LP) only slightly reduced the power of the neoVAN-4S-HP amplifier to \SI{38.2}{\W} without changing the higher-order mode content. It, however, reduced the frequency noise of the seed laser (as presented in Fig.~\ref{fig:frq}) which also improved the laser frequency noise of the amplified beam, as shown in Fig.~\ref{fig:frqneoVAN}.

\begin{figure}[htbp]
        \centering
        \includegraphics[width=\linewidth]{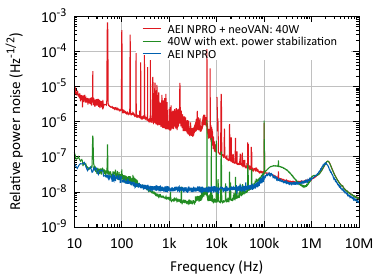}
    \caption{The relative power noise of the amplified beam is shown, along with the demonstration of the noise reduction achieved by implementing a power stabilization loop on the amplifier's pump current. For reference, the relative power noise of the used early version of the AEI NPRO laser is also plotted.
    All shown measurements were performed with engaged internal noise reduction of the AEI NPRO.}
    \label{fig:RPNneoVAN}
\end{figure}

Fig.~\ref{fig:RPNneoVAN} shows a comparison of the relative power noise of the AEI NRPO, of the amplified beam with and without analog feed-back power stabilization via the pump light of the neoVAN amplifier. The uncontrolled neoVAN increases the power noise below \SI{100}{\kHz} whereas the AEI NPRO dominates the free running noise of the amplified beam above this frequency.
As illustrated in Fig.~\ref{fig:setup} a feedback control loop was implemented to stabilize the amplified laser power. The signal of the photodiode (PD$_{\text{il}}$) was processed in an analog servo and its output signal actuated the laser power via the pump current of one neoVAN  pump laser diode. 
Verified with an independent out-of-loop detector (PD$_{\text{ool}}$)
this simple feedback control loop could effectively reduce the relative power noise of the \SI{40}{\W} beam to a level of \SI{1e-8}{\persqrtHz} below \SI{40}{\kHz}, as shown in Fig.~\ref{fig:RPNneoVAN}. The noise increase between \SI{50}{\kHz} and \SI{500}{\kHz} can be mitigated by a more sophisticated control loop design.

\section{Conclusion}

The development of a novel  non-planar ring oscillator (NPRO) lasers operating at a wavelength of \SI{1064}{\nm} has led to a significant enhancement in noise performance compared to previously available models. Our characterizations showed an improvement by more than a factor of 10 in both power noise and frequency noise for the NPRO laser developed in-house within the frequency band of \SI{1}{\Hz} to \SI{10}{\MHz}, compared, for example, to currently employed NPROs in gravitational wave detectors \cite{Kwee2008}. 
The measurements on the AEI NPRO laser presented here, revealed a partial correlation between improvements in power noise and frequency noise, as indicated by the coupling transfer functions from laser power to laser frequency \cite{Harb1997,Heurs2006}. The low-noise performance was achieved by careful design of a low-noise current source, good temperature stabilizations, an internal noise reduction scheme with active control down below \SI{10}{\Hz} and optimization of the spatial overlap between pump and signal field.

We also investigated the compatibility of the AEI NPRO laser with a solid-state amplifier. In combination with a neoVAN 4S-HP amplifier, we demonstrated an excellent performance and achieved output powers of up to \SI{40}{\W} without any impairment of the seed laser frequency noise or beam quality. 
The measured increase in power noise of the amplified beam was mitigated through a feedback control loop that fed back to the pump current in the solid-state amplifier, highlighting the scalability option of our new NPRO without excess noise for high-power applications.

In conclusion, the here reported improved NPRO laser performance can have significant impacts on ground- and space-based gravitational wave detectors \cite{Tse19, Acernese19, Lough21, GeoKagra2022C, Colpi2024}, as well as on geodesy missions like GRACE-missions \cite{Abich2019}. Furthermore, the fields of squeezed light generation ~\cite{Vahlbruch2016, sqz2128nm, LIGOSQL2024}, quantum state engineering~\cite{EPR_ANU, EPR_HH, Virgo_FDS, Baune2016}, particle and molecule trapping and cooling~\cite{DeLosRios2021,Gregory2019}, atomic clock comparison~\cite{Zhang2024}, as well as experiments for searching for new particles, dark matter or quantum gravity signatures~\cite{Kozlowski2025, Patra2024} 
might strongly benefit from lower noise NPROs. 

\begin{acknowledgments}
Funded by the Deutsche Forschungsgemeinschaft (DFG, German Research Foundation) under Germany’s Excellence Strategy – EXC-2123 QuantumFrontiers – 390837967.
\end{acknowledgments}

\section*{Author Declarations}
\subsection*{Conflict of Interest}
The VM Photonics GmbH was founded as a spin-off from the Leibniz University Hannover and Max Planck Institute for Gravitational Physics (AEI Hannover) and now offers the here described NPRO laser commercially.
Henning Vahlbruch and Fabian Meylahn are the owner of this company.
Benno Willke has no conflicts to disclose. 

\subsection*{Author Contributions}

H.V. and F.M. contributed equally to this work.

\noindent\textbf{H. Vahlbruch}: Conceptualization (equal); Investigation (equal); Methodology (equal); Funding
acquisition (lead); Visualization (supporting); Writing - original draft (equal); Writing - review \& editing (equal).
\textbf{F. Meylahn}:
Conceptualization (equal); Investigation (equal); Methodology (equal); Software (lead); Visualization (lead); Writing - original draft (equal); Writing - review \& editing (equal).
\textbf{B. Willke}:
Conceptualization (equal); Funding
acquisition (supporting); Project administration (lead); Supervision (lead); Writing - original draft (supporting); Writing – review \& editing (equal).


\section*{Data Availability Statement}
Data available on request from the corresponding authors.

\bibliography{referencesCollaborations}

\end{document}